\documentclass{PoS}
\usepackage{units}
\usepackage{comment}
\usepackage{graphicx}
\usepackage{subcaption}
\graphicspath{ {./images/} }

\title{A method for Cloud Mapping in the Field of View of the Infra-Red Camera during the EUSO-SPB1 flight}

\ShortTitle{Infra Red Cloud Mapping}

\author{
A.~Bruno$^1$, 
A.~Anzalone$^{1}$, 
\speaker{C.~Vigorito}$^2$, 

for the JEM-EUSO Collaboration \footnote{for collaboration list see PoS(ICRC2019)1177} \\
\llap{$^1$} INAF-IASF Palermo, Istituto di Astrofisica e Fisica Cosmica, Palermo, Italy\\
\llap{$^2$} Dipartimento di Fisica, Universit\`{a} di Torino, Italy

E-mail: \email{alessandro.bruno@inaf.it}
}

\abstract{EUSO-SPB1 was released on April 24\textsuperscript{th}, 2017, from the NASA balloon launch site in Wanaka (New Zealand) and landed on the South Pacific Ocean on May 7\textsuperscript{th}. The data collected by the instruments onboard the balloon were analyzed to search UV pulse signatures of UHECR (Ultra High Energy Cosmic Rays) air showers. Indirect measurements of UHECRs can be affected by cloud presence during nighttime, therefore it is crucial to know the meteorological conditions during the observation period of the detector. During the flight, the onboard EUSO-SPB1 UCIRC camera (University of Chicago Infra-Red Camera), acquired images in the field of view of the UV telescope.
The available nighttime and daytime images, include information on meteorological conditions of the atmosphere observed in two infra-red bands. The presence of clouds has been investigated employing a method developed to provide a dense cloudiness map for each available infra-red image. The final masks are intended to give pixel cloudiness information at the IR-camera pixel resolution that is nearly 4-times higher than the one of the UV-camera. In this work cloudiness maps are obtained by using an expert system based on the analysis of different low-level image features. Furthermore, an image enhancement step was needed to be applied as a preprocessing step to deal with uncalibrated data. 
}

\FullConference{36th International Cosmic Ray Conference -ICRC2019-\\
		July 24th - August 1st, 2019\\
		Madison, WI, U.S.A.}

\begin{document}

\section{Introduction}
 The Joint Experiment Missions for Extreme Universe Space Observatory (JEM-EUSO) is a program for the Ultra High Energy Cosmic Ray (UHECR) observation from space \cite{EUSO-Program}. The program apart from the main missions, includes a set of pathfinders to test  technologies and instruments developed in this framework. With EUSO-TA the sky is observed from the ground with the still operational telescope prototype, while observations from the stratosphere were done  with the past flight of the EUSO-Balloon in 2014, the Super Pressure Balloon EUSO-SPB1 in 2017 and will continue with the future flight of EUSO-SPB2 in 2022, finally from the ISS, with Mini-EUSO this year. Here we focus our attention on the recent flight of EUSO-SPB1. 
 
 EUSO-SPB1 was launched on April 24\textsuperscript{th}, 2017, from the NASA site in Wanaka (New Zealand) and landed on the South Pacific Ocean on May 7\textsuperscript{th}. The flight was unexpectedly interrupted due to a sudden problem occurred to the balloon envelope. Despite the short duration, the devices for UHECR detection, and  the ancillary sensor devoted to monitor the atmosphere from the meteorological point of view, could be tested and were able to take data. One of the main concern is the presence of clouds in the UV-telescope Field of View (FoV) that can affect UV data acquisition and UHECR energy/direction reconstruction phases. During its nearly 12 day flight, a small number of images was acquired by the UCIRC camera (University of Chicago Infra-Red Camera) \cite{Wiencke}, with the objective of providing a set of data of the various cloudy scenes seen during the observation period of the UV detector and in daytime. The IR system consisted of two sensors that operated in two IR bands centred at 10.5$\mu$m and 12$\mu$m, with a $\sim$32$^\circ$x24$^\circ$ FoV, that fully covered the 11$^\circ$x11$^\circ$ FoV of the UV telescope. At regular time intervals, they were expected to take images of the atmosphere below the balloon in such way cloud top temperatures would have been retrieved by analyzing the calibrated data in both bands. The camera system sometime did not work as expected, however a quantity of data that were not calibrated and shoot mostly in daytime, are available. In general starting from cloud temperature data, several methods can be used \cite{Anzalone},\cite{Kenji} to reconstruct the cloud coverage information and the cloud top height of the cloudy pixels. In the following sections we present the preliminary results of a method applied to the available images, to map the cloudiness in the IR camera FoV, despite several issues affecting the data.

 \section{Proposed Method}
 In this section, a more detailed description of our proposed method for Cloud Mapping in the Field of View of the Infra-Red Camera during the EUSO-SPB1 flight is given. The presence of clouds has been investigated employing a method developed to provide a dense cloudiness map for each available infra-red image. In greater detail, cloudiness maps are extracted with a three-module system that consists of a Deep Learning architecture, and a low-level image feature analysis 
 The first module of our system is based on Deep Learning CNN (Convolutional Neural Network) \cite{Lecun} technique, which is able to classify image patches as cloudy or clear, and the accurate Data Enhancement and Cloud Mapping modules. Some considerations must be mentioned before going on describing the proposed method. The images that have been treated in this work suffer from noise because of some calibration problems, this is why it has been necessary to treat them with some image enhancement procedures to improve the overall performances of the proposed method. In the next subsections more details will be given to enrich the description of the following steps: Image Classification; Data Enhancement; Cloud Mapping. 

\subsection{Image Classification}
\begin{figure}
    \centering
    \includegraphics[scale=0.5]{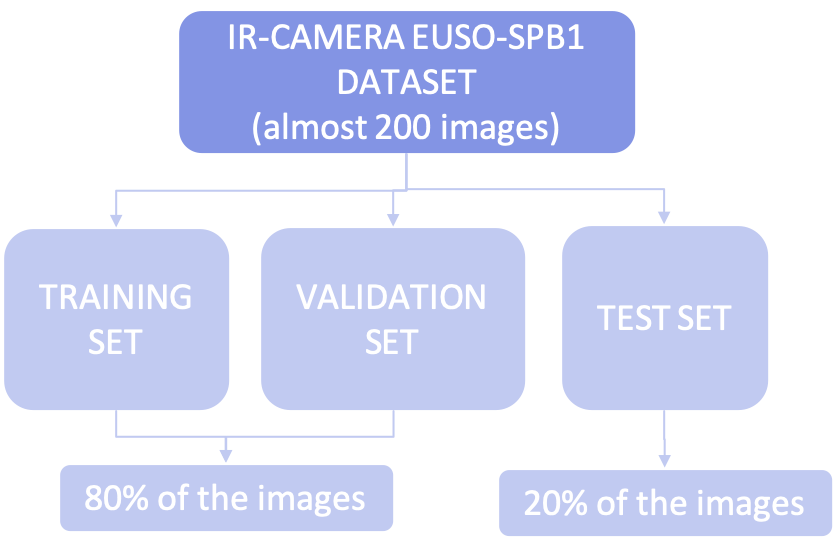}
    \caption{Dataset used in Deep Learning module}
    \label{fig:deeplearning1}
\end{figure}

\begin{figure}
    \centering
    \includegraphics[scale=0.5]{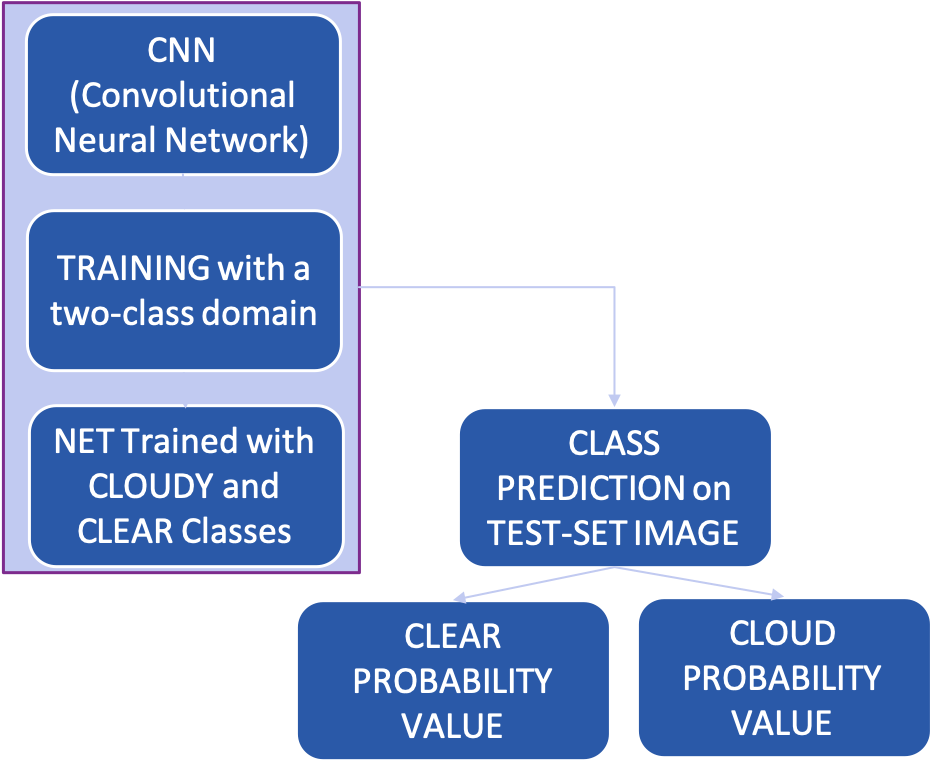}
    \caption{Deep Learning Flow Chart}
    \label{fig:deeplearning2}
\end{figure}

The Image Classification module is based on Deep Learning paradigm. CNNs (Convolutional Neural Networks) are employed because of their high accurate rates in tasks as image and object classification \cite{Lecun} \cite{GoogleNet}. As for most of the machine learning methods labelled samples are needed to train the model, which is then, able to retrieve knowledge using data inference. The Deep Learning module is employed only to classify cloudy and clear images. The pixel-wise cloudiness investigation is demanded to the cloud mapping module that is mainly based on low-level image processing techniques and will be described in the next subsection. To set-up a Deep Learning solution for the purpose of discriminating cloudy images from clear images, a new dataset with infra-red images coming from EUSO-SPB1 is needed. Almost 200 images are chosen and divided with respect to the cloudiness coverage, 80\% of images are used as training and validation set and the rest 20\% is used as test set (figure  \ref{fig:deeplearning1}). As it can be seen in figure \ref{fig:deeplearning2}, the classification task is conceived as a two-class problem (clear and cloudy classes). Once the training model is built the effectiveness of the model is assessed over new data (test set). 
The good thing of finding a solution based on Deep Learning paradigm is that it seems to be quite robust against noise from images. Several experimental sessions on both noisy and noiseless image samples show quite accurate ability of CNN to predict cloudy images from noisy images despite the high level of noise embedded in the images themselves. 
In greater detail, GoogleNet \cite{GoogleNet} is adopted as pre-trained Convolutional Neural Network to be fine-tuned with a further training step (called Transfer Learning \cite{Transfer_Learning})  over EUSO-SPB1 infra-red images. The aforementioned step passes for Transfer Learning because it is employed to transfer the knowledge of networks such as GoogleNet, which are pre-trained over ImageNet database \cite{ImageNet} over new data coming from new domains. The Deep Learning module gives out probability values of data be falling within the training classes. This is because the last layer of CNN is a Softmax function, which shows probability values of test data to fall withing the problem classes. 
In this work it is adopted an experimentally assessed threshold to decide when a patch is cloudy or clear. 

\subsection{Image Enhancement}
\begin{figure}
\centering
\begin{subfigure}{0.5\textwidth}
  \centering
  \includegraphics[width=0.5\textwidth]{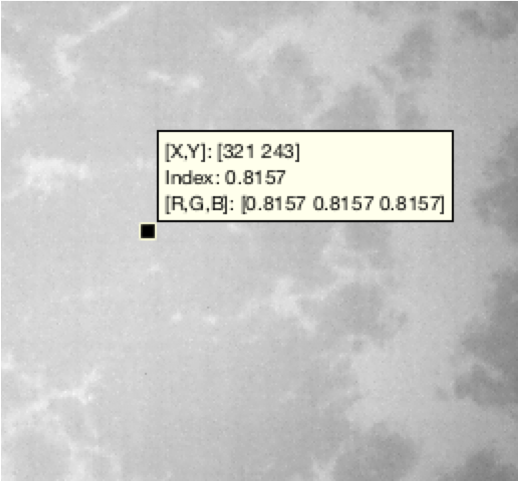}
  \caption{}
  \label{fig:noiseA}
\end{subfigure}%
\begin{subfigure}{0.5\textwidth}
  \centering
  \includegraphics[width=0.5\textwidth]{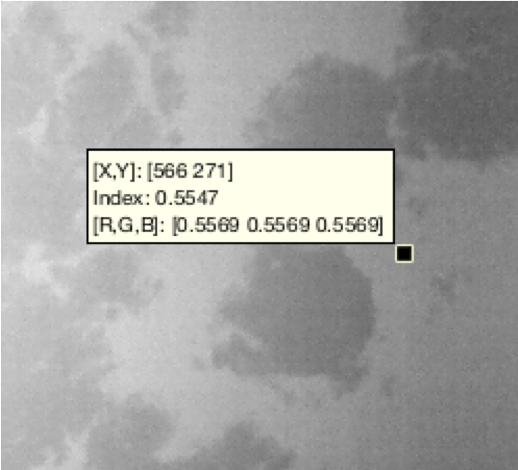}
  \caption{}
  \label{fig:noiseB}
\end{subfigure}
\caption{Some examples of images affected by noise}
\label{fig:noise}
\end{figure}

Due to the missing conversion of the raw data into Brightness Temperature, some Image Enhancement techniques are needed to make infra-red images showing pixel intensity values more coherent with respect to the uncalibrated ADC counts. As briefly mentioned in the previous sections, images from EUSO-SPB1 infra-red camera suffer from calibration problems that affect each image with additive and convolution noise. Looking at figure \ref{fig:noise} it is noticeable some peculiar values of pixel intensity with respect to macro-category objects in our images: cloudy pixels are generally expected to show intensity values (in the infra-red domain) lower than cloud-free pixels falling on sea. Therefore, dark regions in the image may correspond to clouds whilst the bright regions may correspond to clear sky, then it is expected that pixels belonging to dark area show lower intensity values than the light area pixels along different images. Unfortunately, due to the aforementioned noise, this is not (see figures \ref{fig:noiseA} and \ref{fig:noiseB}) what has been observed in our infra-red images, where there are many pixels belonging to different image regions that behave not coherently.

\begin{figure}
\centering
\begin{subfigure}{0.5\textwidth}
  \centering
  \includegraphics[width=0.5\textwidth]{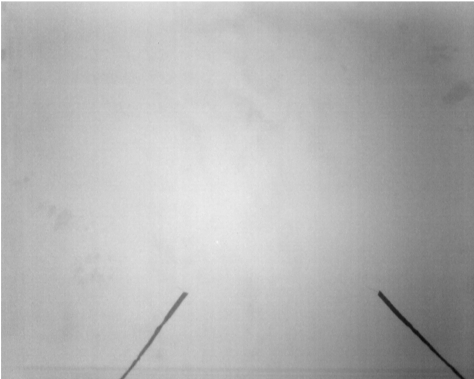}
  \caption{}
  \label{fig:stripesA}
\end{subfigure}%
\begin{subfigure}{0.5\textwidth}
  \centering
  \includegraphics[width=0.5\textwidth]{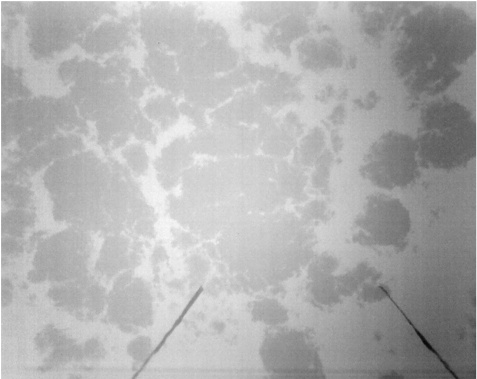}
  \caption{}
  \label{fig:stripesB}
\end{subfigure}
\caption{Non negligible number of pixels are cluttered by floating strips (dark objects in the bottom part of the images (a-b))}
\label{fig:stripes}
\end{figure}

Furthermore, as it can be seen in figure \ref{fig:stripes}, a lot of pixels are cluttered because of the presence of floating strips from the balloon. 

Therefore two techniques are necessary to enhance the image readability: image denoising to treat the noise present all over the image; image restoration to eliminate the floating strips from the scene. 
To deal with noise, it is needed to analyze what kind of noise affects the image. 

\begin{figure}
\centering
\begin{subfigure}{0.33\textwidth}
  \centering
  \includegraphics[width=0.7\textwidth]{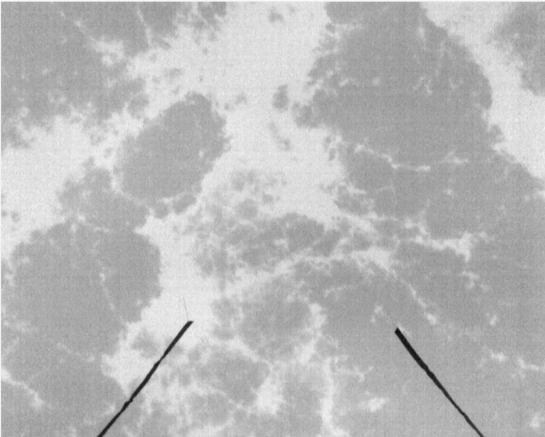}
  \caption{}
  \label{fig:denoiseA}
\end{subfigure}%
\begin{subfigure}{0.33\textwidth}
  \centering
  \includegraphics[width=0.7\textwidth]{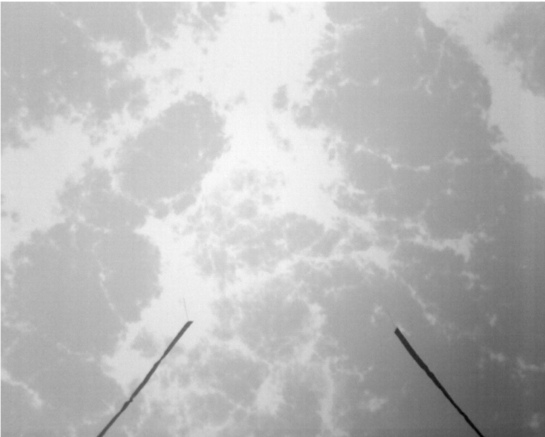}
  \caption{}
  \label{fig:denoiseB}
\end{subfigure}
\begin{subfigure}{0.33\textwidth}
  \centering
  \includegraphics[width=0.7\textwidth]{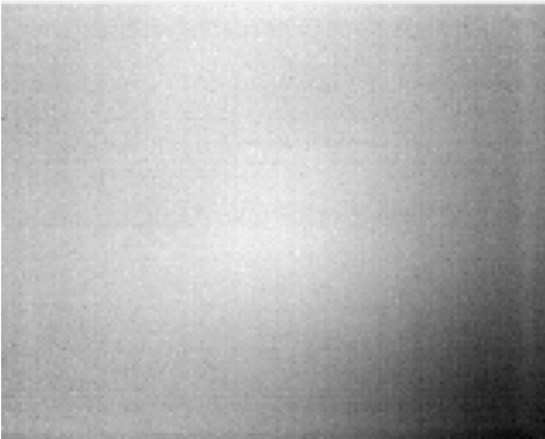}
  \caption{}
  \label{fig:denoiseC}
\end{subfigure}
\caption{The result of equation \ref{eq:1} shows an image with pixel values that are less affected by noise with respect to the original image (b). This has been performed through the analysis of the noise in homogeneous patches like the one in (c)} 
\label{fig:denoise}
\end{figure}

The canonical image restoration approach as in equation \ref{eq:1}, is applied to analyse the noisy image g(x,y) and its own relation to noiseless signal f(x,y) and the noise signals h(x,y) and n(x,y).
We assumed h(x,y)=1 and noise n(x,y)=c*n\textsubscript{0}(x,y),
where the noise distribution n\textsubscript{0}(x,y), was retrieved from the analysis of the noise in homogeneous patches like the one in figure \ref{fig:denoiseC}. c is a scaling factor which was determined experimentally.
\begin{equation}
    g(x,y) = h(x,y) * f(x,y) + n(x,y)
    \label{eq:1}
\end{equation}



To retrieve the noise model distribution, an image from EUSO-SPB1 on 2017 April 20\textsuperscript{th} is picked up, EUSO-SPB1 did not take off yet and the surface taken with the infra-red camera was quite homogeneous (see figure \ref{fig:stripesA}). Image denoising results are given as in figure \ref{fig:denoise}.   
As it can be observed in the figure, the image that undergoes the denoising step 
shows pixel values that are more coherent across different regions and locations in the images (the centered bias effect is lower than in the original image and the  underexposure effect along the image borders is quite diminished).

\begin{figure}
\centering
\begin{subfigure}{0.33\textwidth}
  \centering
  \includegraphics[width=0.7\textwidth]{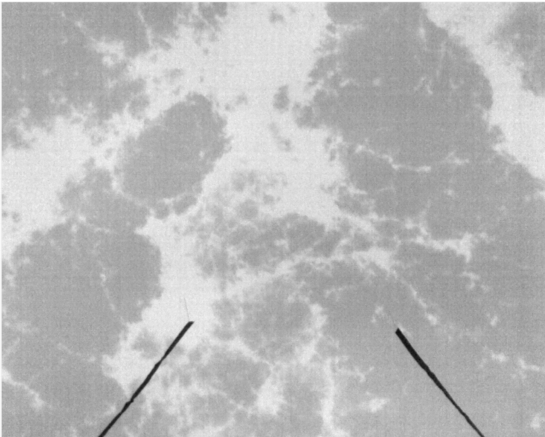}
  \caption{}
  \label{fig:restorationA}
\end{subfigure}%
\begin{subfigure}{0.33\textwidth}
  \centering
  \includegraphics[width=0.7\textwidth]{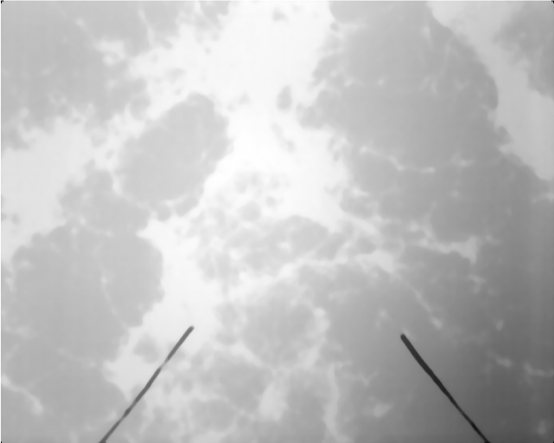}
  \caption{}
  \label{fig:restorationB}
\end{subfigure}
\begin{subfigure}{0.33\textwidth}
  \centering
  \includegraphics[width=0.7\textwidth]{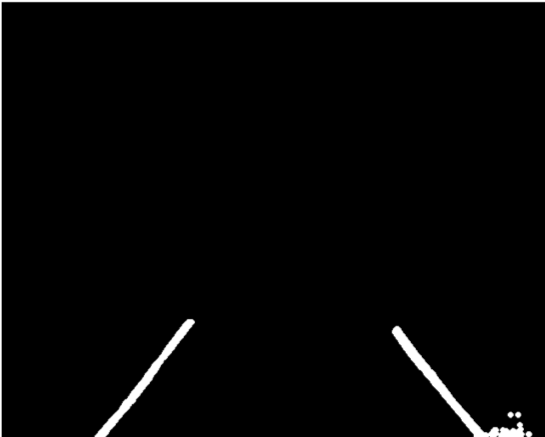}
  \caption{}
  \label{fig:restorationC}
\end{subfigure}
\begin{subfigure}{0.33\textwidth}
  \centering
  \includegraphics[width=0.7\textwidth]{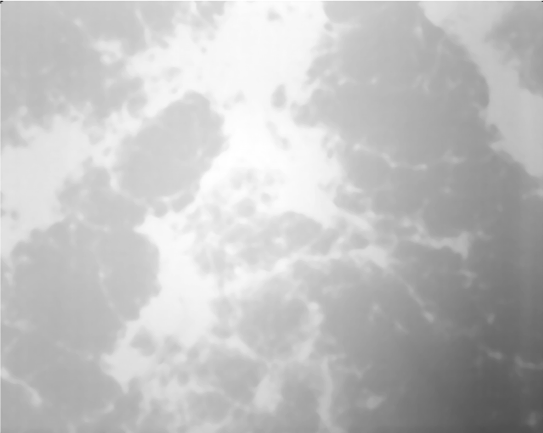}
  \caption{}
  \label{fig:restorationD}
\end{subfigure}

\caption{Strip removal process. Denoised image (a). Smoothed image by 7x7 median filter (b). Strip binary mask (c). Restored image (d)} 
\label{fig:restoration}
\end{figure}

As far as it concerns the image restoration approach to remove the floating strips, a 7 x 7 Median Filter Mask applied to the denoised image (see figure \ref{fig:restorationA}) allows for achieving an image with smoother strip regions (see figure \ref{fig:restorationB}) characterized by a particular gray-level value range. Then a Morphological Operator (dilation) is applied to the image to make the floating strips thicker giving out a binary mask (see figure \ref{fig:restorationC}). The image is then restored using the floating strip binary mask and filling out the white pixels with the neighborhood values (see figures \ref{fig:restorationD}).

\subsection{Cloud Mapping}
\begin{figure}
\centering
\begin{subfigure}{0.5\textwidth}
  \centering
  \includegraphics[width=0.5\textwidth]{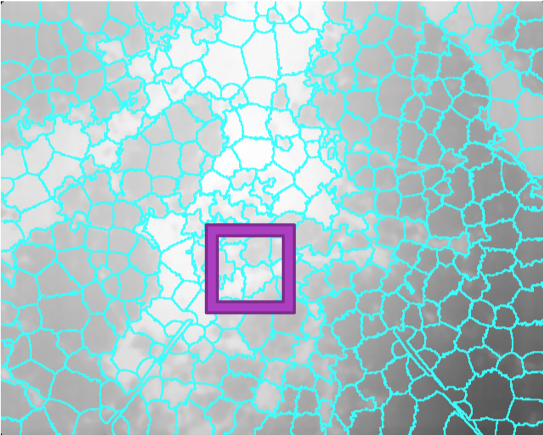}
  \caption{}
  \label{fig:clustering1}
\end{subfigure}%
\begin{subfigure}{0.5\textwidth}
  \centering
  \includegraphics[width=0.44\textwidth]{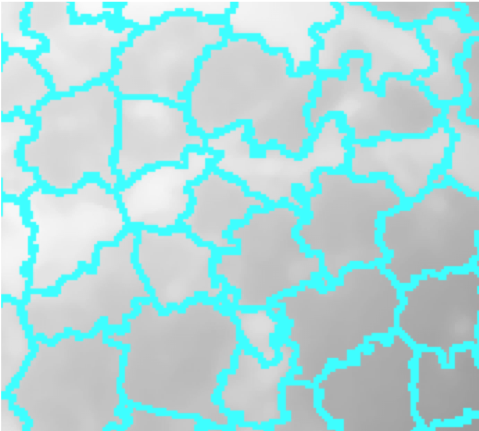}
  \caption{}
  \label{fig:clustering2}
\end{subfigure}
\caption{SLIC is applied to the enhanced image to collect superpixels (a). Pixels falling within same clusters share common properties such as similar positions and intensity levels (b)}
\label{fig:clustering}
\end{figure}

\begin{figure}
\centering
\begin{subfigure}{0.33\textwidth}
  \centering
  \includegraphics[width=0.7\textwidth]{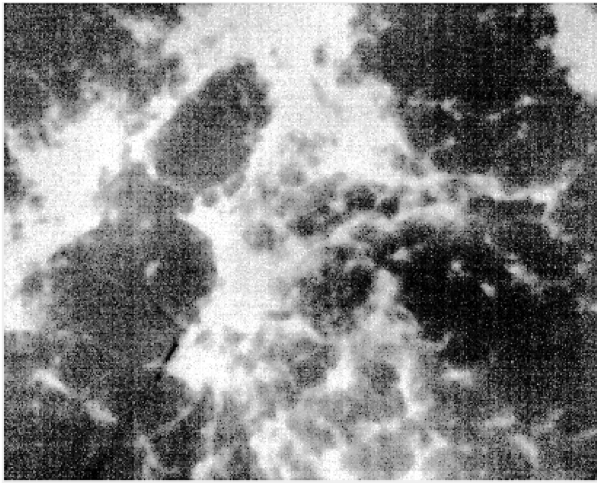}
  \caption{}
  \label{fig:histogramA}
\end{subfigure}%
\begin{subfigure}{0.33\textwidth}
  \centering
  \includegraphics[width=0.7\textwidth]{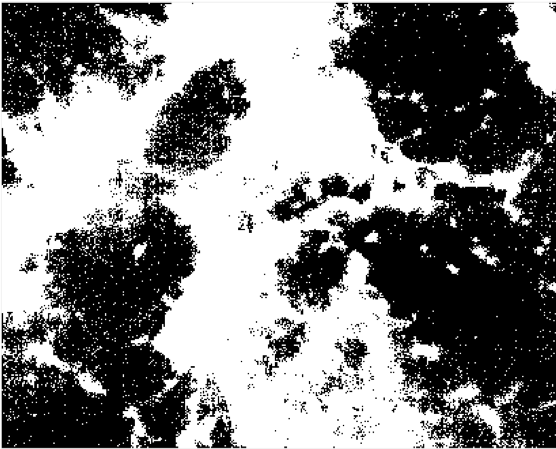}
  \caption{}
  \label{fig:histogramB}
\end{subfigure}
\caption{Hisogram Equalization is needed (a) to stretch out pixel values along two main modalities (cloudy, clear pixels), a simple threshold based binarization allows for cloud mask (b)} 
\label{fig:histogram}
\end{figure}

In this section several steps are discussed: pixel clustering; histogram equalization, binarization and post-processing. In this context binarization is meant to be the process that makes an image binary, each pixel can assume either 0 or 1 value. 
As first step of this module pixel clustering is applied to collect superpixels sharing common properties such as intensity value and location in the image, the very popular SLIC \cite{SLIC} algorithm is adopted as clustering method.  It is shown that local key-points are extracted along the borders of the clouds in the image. Starting from this consideration a regional approach such as region growing is applied to connect all those neighbour pixels showing visual properties of interest for our purpose. SLIC needs the number of clusters as parameter to go for the separation of pixels with respect to gray level intensity and location; this value is experimentally fixed. As it can be seen in figure \ref{fig:clustering2}, pixels falling within each cluster show approximately quite similar gray-level value. 
Histogram equalization is needed to increase the image contrast (see figure \ref{fig:histogramA}), which helps finding a clearer pixel distribution with respect to cloudy and clear classes. 
A simple threshold based step gives out, as result, a binary image (each pixel can assume either 0 or 1 value): the SLIC map (in the same way as in figure \ref{fig:clustering1}) is applied to the output of the equalized image that is filtered with a fixed threshold (0.5) (see figure \ref{fig:histogramB}). The final mask is intended to give pixel cloudiness information at the IR-camera pixel resolution that is nearly 4-times higher than the one of the UV-camera. So that the future step will be mapping the IR-camera FoV with the UV-telescope FoV, in order to make useful the retrieved cloudiness masks.

\section{Experimental Results}
Since the proposed method is basically an integrated solution for the cloud mapping from EUSO-SPB1 infra-red images, it is needed to describe the experimental set-up for each module.
All the images processed in this work are with size of 512 x 640 pixels and are normalised in the range [0,1]. Starting by analysing the Deep Learning module, it is important to mention that GoogleNet is adopted as CNN, the training step involves almost 200 images with size 512 x 640 and it takes 2 hours while the running time to classify an image is about real-time. Steps such as image denoising and image restoration are handled with MATLAB, which makes matrix processing (image filtering, image filling, histogram equalization) quite straightforward and computationally lightweight (just few milliseconds per operation), the most time-consuming image processing step is the pixel clustering (SLIC algorithm takes 0.35 seconds per each image to divide all pixels with 200 clusters, which is experimentally fixed). The last experimental considerations highlight that, apart from the training step, all the processing steps of the proposed solution are lightweight and easy to perform, they can be run in workstations equipped with old-fashion non-efficient components as well.

\section{Conclusions and Future Works}
The main objective of this work is to employ different image processing and computer vision techniques to extract as much information as possible from images that suffer from different constraints and problems such as pixels cluttered by objects (floating strips), noise across centre and borders of images caused by missing calibration. As described in the previous sections, CNNs are employed to accomplish the Image Classification step while Data Enhancement operations are needed for the Cloud Mapping to make sufficiently working.
The performances of CNN show how much robust is the Deep Learning approach against the noise distribution all over the images from EUSO-SPB1. The image denoising step is quite successful in the aim of normalizing all pixel values and subtracting the additive noise from them. The image restoration allows for erasing floating strip pixels and filling them out with nearest neighbour values. The cloud mapping is eventually obtained  with a threshold based binarization of the equalized image clusters obtained with the SLIC algorithm. To assess the accuracy percentage of the cloud mapping it would be necessary the corresponding ground-truth manually generated that is not available for the images in this work.
However, the preliminary results of the proposed method are quite encouraging and prompt to keep going on with this approach using some other techniques such as Transfer Learning and Semantic Segmentation to assess and evaluate the goodness of our cloud masks. Any chance to have the images manually labelled would come out as a new challenge to implement, develop and test new machine learning and deep learning techniques for the cloud mapping that would be probably running with no needs of other further steps (as currently arranged).  

\section{Acknowledgements}
This work was partially supported by Basic Science Interdisciplinary Research Projects of RIKEN and JSPS KAKENHI Grant (22340063, 23340081, and 24244042), by the Italian Ministry of Foreign Affairs and International Cooperation, by the Italian Space Agency through the ASI INFN agreement n. 2017-8-H.0, by contract contract 2016-1-U.0, by NASA award 11-APRA-0058 in the USA, by the Deutsches Zentrum f\"ur Luft- und Raumfahrt, by the French space agency CNES, the Helmholtz Alliance for Astroparticle Physics funded by the Initiative and Networking Fund of the Helmholtz Association (Germany), and by Slovak Academy of Sciences MVTS JEMEUSO as well as VEGA grant agency project 2/0132/17. Russia is supported by ROSCOSMOS and the Russian Foundation for Basic Research Grant No 16-29-13065. Sweden is funded by the Olle Engkvist Byggm\"astare Foundation.

\end{document}